\documentclass[preprint,preprintnumbers,nofootinbib,tightenlines,12pt,superscriptaddress]{revtex4}
\pdfoutput=1

\RequirePackage[12tabu, orthodox]{nag}
\usepackage[all, warning]{onlyamsmath}

\usepackage{amsmath,amssymb,setspace,amsfonts,latexsym,braket}
\usepackage[colorlinks]{hyperref}
\usepackage{graphicx}
\usepackage{color}

\newcommand\unit[1]{\,{\rm #1}}

\newcommand\keV{\unit{keV}}
\newcommand\MeV{\unit{MeV}}
\newcommand\GeV{\unit{GeV}}

\newcommand{\hc}{{\rm h.c.}}
\newcommand\vev[1]{\left\langle #1 \right\rangle}
\newcommand\Order{{\cal O}}
\newcommand\bcm[3]{\left(\frac{#1}{#2}\right)^{#3}}

\begin{document}

\hfill CTPU-PTC-18-10

\hfill UMN-TH-3716/18

\hfill FTPI-MINN-18/07

\hfill UT-18-10

\title{\Large Self-interacting dark matter and muon $(g-2)$ in a gauged U$(1)_{L_{\mu} - L_{\tau}}$ model}
\author{Ayuki Kamada} 
\affiliation{Center for Theoretical Physics of the Universe, Institute for Basic Science (IBS), Daejeon 34126, Korea}
\author{Kunio Kaneta}
\affiliation{Center for Theoretical Physics of the Universe, Institute for Basic Science (IBS), Daejeon 34126, Korea}
\affiliation{School of Physics and Astronomy, University of Minnesota, Minneapolis, MN 55455, USA}
\affiliation{William I. Fine Theoretical Physics Institute, University of Minnesota, Minneapolis, MN 55455, USA}
\author{Keisuke Yanagi}
\affiliation{Department of Physics, University of Tokyo, Bunkyo-ku, Tokyo 113-0033, Japan}
\author{Hai-Bo Yu}
\affiliation{Department of Physics and Astronomy,
University of California, Riverside, California 92521, USA}


\begin{abstract}

We construct a self-interacting dark matter model that could simultaneously explain the observed muon anomalous magnetic moment. It is based on a gauged U$(1)_{L_{\mu} - L_{\tau}}$ extension of the standard model, where we introduce a vector-like pair of fermions as the dark matter candidate and a new Higgs boson to break the symmetry. The new gauge boson has a sizable contribution to muon $(g-2)$, while being consistent with other experimental constraints. The U$(1)_{L_{\mu} - L_{\tau}}$ Higgs boson acts as a light force carrier, mediating dark matter self-interactions with a velocity-dependent cross section. It is large enough in galaxies to thermalize the inner halo and explain the diverse rotation curves and diminishes towards galaxy clusters. Since the light mediator dominantly decays to the U$(1)_{L_{\mu} - L_{\tau}}$ gauge boson and neutrinos, the astrophysical and cosmological constraints are weak. We study the thermal evolution of the model in the early Universe and derive a lower bound on the gauge boson mass. 

\end{abstract}

\maketitle



\newpage

\section{Introduction}
\label{sec:intro}

Dark matter (DM) makes up $85\%$ of the mass in the Universe, but its nature remains largely unknown.
There has been great progress in studying particle DM candidates associated with extensions of the standard model (SM) of particle physics, which can be probed in high-energy and intensity-frontier terrestrial experiments (see, e.g., Ref.~\cite{Battaglieri:2017aum}). Astrophysical and cosmological observations can also provide important clues to the nature of DM. In fact, a number of astrophysical observations indicate that the cold dark matter (CDM)  model may break down on galactic scales (see Refs.~\cite{Tulin:2017ara, Bullock:2017xww} for a review), although it works remarkably well in explaining large-scale structure of the Universe, from Mpc to Gpc scales. For example, the galactic rotation curves of spiral galaxies exhibit a great diversity in inner shape~\cite{Oman:2015xda}, which is hard to understand in CDM. It has been shown that the diverse rotation curves can be explained naturally if DM has strong self-interactions~\cite{Kamada:2016euw, Creasey:2016jaq}. The analysis has been extended to the dwarf spheroidal galaxies in the Milky Way~\cite{Valli:2017ktb} and other galactic systems~\cite{Kaplinghat:2015aga, Bondarenko:2017rfu}. This self-interacting dark matter (SIDM) scenario has rich implications for understanding the stellar kinematics of dwarf spheroidal galaxies and galaxy clusters and interpreting DM direct, indirect, and collider search results (see, e.g., Ref.~\cite{Tulin:2017ara}).


On the particle physics side, there is also a long-standing puzzle that the measured muon anomalous magnetic moment, $(g-2)_\mu$, is larger than predicted in the SM at the $3\sigma$ level~\cite{Hagiwara:2011af,Davier:2010nc}. This discrepancy may indicate that there is new physics beyond the SM associated with the muon sector. In this work, we propose a model that could simultaneously explain the $(g-2)_\mu$ measurement and provide a realization of SIDM. It is based on a gauged U$(1)_{L_{\mu} - L_{\tau}}$ extension of the SM (see, e.g., Refs.~\cite{He:1991qd, He:1990pn}). Aside from the new gauge ($Z^{\prime}$) and Higgs ($\varphi$) bosons related to the symmetry, we introduce a vector-like pair of fermions ($N$ and $\bar{N}$) as the DM candidate and assume that they couple to $\varphi$ via a Yukawa interaction. The model overcomes a number of challenges in SIDM model building~\cite{Kaplinghat:2013yxa} (see also, e.g, Refs.~\cite{CyrRacine:2012fz, Cline:2013pca,Boddy:2014yra,Boddy:2014qxa,Bernal:2015ova, Boddy:2016bbu, Bringmann:2016ilk,Binder:2017lkj,Ma:2017ucp,Huo:2017vef,Baldes:2017gzu,Bringmann:2018jpr,Zhu:2018tzm,Duerr:2018mbd}). Our main observations are the following:
\begin{itemize}
\item{The presence of $Z^{\prime}$ could contribute to $(g-2)_\mu$. If $Z^{\prime}$ has a mass in the range of $m_{Z'}\sim10\textup{--}100~{\rm MeV}$, there is a viable parameter space to explain the $(g-2)_\mu$ anomaly.}
\item{Astrophysical observations indicate that the DM self-scattering cross section per unit mass is $\sigma / m \gtrsim 1 \unit{cm^{2} / g}$ in galaxies, while diminishing to $\sigma / m \lesssim 0.1 \unit{cm^{2} / g}$ in galaxy clusters~\cite{Kaplinghat:2015aga}. The desired velocity-dependence of $\sigma/m$ can be achieved if $m_\varphi\sim10\textup{--}100~{\rm MeV}$}.
\item{In the early Universe, the mediator $\varphi$ is in equilibrium with $Z^{\prime}$ and the SM neutrinos ($\nu_\mu$ and $\nu_\tau$), and its number density is Boltzmann suppressed when the temperature is below its mass. Thus, the model avoids over-closing the Universe. }
\item{Since $Z^{\prime}$ and $\varphi$ dominantly decay to the neutrinos, the big bang nucleosynthesis (BBN), cosmic microwave background (CMB), and indirect detection bounds are significantly weaker, compared to the models with electrically charged final states~\cite{Kaplinghat:2015gha,Bringmann:2016din}.}
\end{itemize}

This paper is organized as follows. We introduce the U$(1)_{L_{\mu} - L_{\tau}}$ model and discuss relevant experimental and cosmological constraints in Sec.~\ref{sec:model}. We discuss DM phenomenology of the model in Sec.~\ref{sec:dmpheno} and devote Sec.~\ref{sec:concl} to conclusions. In the appendix, we provide detailed discussion on the neutrino-electron scattering cross section induced by $Z'$ (A), the temperature evolution of the model in the early Universe (B), and the decay width of $\varphi$ (C).

\section{The Model}
\label{sec:model}

We introduce a vector-like pair of fermions $N$ and $\bar N$ and a U(1)$_{L_{\mu} - L_{\tau}}$ Higgs $\Phi$ in addition to the U(1)$_{L_{\mu} - L_{\tau}}$ gauge boson $Z_{\mu}^{\prime}$.
The charge assignments are summarized in Table~\ref{tab:quantum_number}.
\begin{table}[h]
  \centering
  \begin{tabular}{ccccc}\hline
            & & SU(2) & U(1)$_{Y}$ & U(1)$_{L_{\mu} - L_{\tau}}$ \\\hline
    fermion & $L_{2} = (\nu_{\mu},\ \mu_L)$ & $2$ & $-1/2$ & $+1$\\
     & $L_{3} = (\nu_{\tau},\ \tau_L)$ & $2$ & $-1/2$ & $-1$\\
     & ${\bar \mu} = \mu_{R}^{\dagger}$ & $1$ & $+1$ & $-1$\\
     & ${\bar \tau} = \tau_{R}^{\dagger}$ & $1$ & $+1$ & $+1$\\
     & $N$ & $1$ & $0$ & $1/2$ \\
     & ${\bar N}$ & $1$ & $0$ & $-1/2$ \\\hline
    scalar &  $H$ (SM Higgs) & $2$ & $1/2$ & $0$ \\ 
     &  $\Phi$ & $1$ & $0$ & $-1$ \\ \hline
  \end{tabular}
  \caption{Charge assignments of the particles in the model, where the fermions are represented by the left-handed 2-component Weyl spinor.
  }
  \label{tab:quantum_number}
\end{table}

The renormalizable Lagrangian density can be written as
\begin{align}
  \label{eq:lag1}
  {\cal L}
   =& {\cal L}_{\rm SM}
    + g^{\prime} Z_{\mu}^{\prime}
    \left( L_{2}^{\dagger} {\bar \sigma}^{\mu} L_{2} - L_{3}^{\dagger} {\bar \sigma}^{\mu} L_{3}
    - {\bar \mu}^{\dagger} \bar{\sigma}^{\mu} {\bar \mu} + {\bar \tau}^{\dagger} \bar{\sigma}^{\mu} {\bar \tau} \right)
    - \frac{1}{4} Z_{\mu \nu}^{\prime} Z^{\prime \mu \nu}
    - \frac{1}{2} \epsilon \, Z_{\mu\nu}^{\prime} B^{\mu \nu}
    \notag \\
   &+ (D_{\mu} \Phi)^{\dagger} D^{\mu} \Phi - V(\Phi, H)
    + i N^{\dagger} {\bar \sigma}^{\mu} D_{\mu} N
    + i {\bar N}^{\dagger} \bar{\sigma}^{\mu} D_{\mu} {\bar N}
    \notag \\
   &- m_{N} N \bar{N}
    - \frac{1}{2} y_{N} \Phi NN
    - \frac{1}{2} y_{\bar N} \Phi^{*} \bar{N} \bar{N} + \hc
\end{align}
The covariant derivative on U(1)$_{L_{\mu} - L_{\tau}}$-charged particles is written as $D_{\mu} = \partial_{\mu} - i g^{\prime} Q Z^{\prime}_{\mu}$, where $Q$ is a U(1)$_{L_{\mu} - L_{\tau}}$ charge.
The field strengths of $Z^{\prime}_{\mu}$ and the U(1)$_{Y}$ gauge boson $B_{\mu}$ are denoted by $Z^{\prime}_{\mu \nu}$ and $B_{\mu \nu}$, respectively.
The scalar potential of $\Phi$ takes a form of 
\begin{align}
  \label{eq:potential}
  V(\Phi, H)
  =
  - m_{\Phi}^{2} |\Phi|^{2}
  + \frac{1}{4} \lambda_{\Phi} |\Phi|^{4}
  + \lambda_{\Phi H} |\Phi|^{2} |H|^{2} \,.
\end{align}
We set $\lambda_{\Phi H} = 0$ so that it does not induce the DM interaction with SM particles through the Higgs portal.
$\Phi$ develops a vacuum expectation value (VEV) and can be expanded as $\Phi(x) = (v_{\Phi} + \varphi(x)) / \sqrt{2}$ in the unitary gauge, where $v_{\Phi} \simeq 2 \sqrt{m_{\Phi}^{2} / \lambda_{\Phi}}$.
The resultant masses of $Z^{\prime}$ and $\varphi$ are given by $m_{Z^{\prime}} = g^{\prime} v_{\Phi}$ and $m_{\varphi} = \sqrt{\lambda_{\Phi} / 2} \, v_{\Phi}$, respectively.
Note that the VEV of $\Phi$ breaks U(1)$_{L_{\mu} - L_{\tau}}$ into a Z$_{2}$ symmetry, which stabilizes the lightest state of $N$ and $\bar N$, i.e., the DM candidate.
In our model, $\varphi$ has a mass of $\Order(10) \MeV$ and plays a role of the SIDM mediator.

We assume that there is no kinetic mixing between $Z^{\prime}$ and $B$ at some high-energy scale, $\epsilon=0$.
This can be achieved by imposing a charge conjugation symmetry $C_{L_{\mu} - L_{\tau}}$: $L_{2} \leftrightarrow L_{3}$, $\mu \leftrightarrow \tau$, $N \leftrightarrow {\bar N}$, $Z^{\prime} \to - Z^{\prime}$, and $\Phi \to \Phi^{*}$.
However, since this symmetry is broken by the Yukawa mass terms of $\mu$ and $\tau$, the mixings of $Z^{\prime}$ with the photon $A$ and with the $Z$ boson arise at the 1-loop level at low energy.
The $A$-$Z^{\prime}$ mixing below $m_{\mu}$ is
\begin{align}
\label{eq:azp-mix}
\epsilon_{A Z^{\prime}} = - \frac{e g^{\prime}}{12 \pi^{2}} \ln \left( \frac{m_{\tau}^{2}}{m_{\mu}^{2}} \right) \,,
\end{align}
where $e$ is the electric charge of the electron and $m_{\mu}$ and $m_{\tau}$ are the mu and tau lepton masses, respectively.
To obtain the canonical gauge fields, we redefine the field as $A \to A + \epsilon_{A Z^{\prime}} Z^{\prime}$ for $|\epsilon_{A Z^{\prime}}| \ll 1$.
It induces a coupling between $Z^{\prime}$ and the SM electromagnetic current:
\begin{equation}
  \label{eq:zp-q}
  {\cal L}_{Z^{\prime}  {\rm em}}
  =
  - \epsilon_{A Z^{\prime}} e Z^{\prime}_{\mu} J^{\mu}_{\rm em} \,.
\end{equation}

While the $Z$-$Z^{\prime}$ mixing is given by
\begin{align}
\epsilon_{Z Z^{\prime}} = \left(- \frac{1}{4} + s_{W}^{2} \right) \frac{e g^{\prime}}{12 \pi^{2} s_{W} c_{W}} \ln \left( \frac{m_{\tau}^{2}}{m_{\mu}^{2}} \right) \,,
\end{align}
where $c_{W} = \cos \theta_{W}$ and $s_{W} = \sin \theta_{W}$ with $\theta_{W}$ being the Weinberg angle.
After performing the following field shifts: $Z \to Z - \epsilon_{Z Z^{\prime}} r_{Z Z^{\prime}}^{2} Z^{\prime}$ and $Z^{\prime} \to Z^{\prime} + \epsilon_{Z Z^{\prime}} Z$, for $|\epsilon_{Z Z^{\prime}}| \ll r_{Z Z^{\prime}} \equiv m_{Z^{\prime}}/m_{Z} \ll 1$, we have the coupling of $Z$ to the U(1)$_{L_{\mu} - L_{\tau}}$ current,
\begin{equation}
  {\cal L}_{L_{\mu} - L_{\tau} Z}
  =
  - g^{\prime} \epsilon_{Z Z^{\prime}} Z_{\mu} J_{L_{\mu} - L_{\tau}}^{\mu} \,,
\end{equation}
and the coupling of $Z^{\prime}$ to the SM neutral current,
\begin{equation}
  \label{eq:zp-z}
  {\cal L}_{Z^{\prime} {\rm neu}}
   =
  \frac{e}{s_{W} c_{W}} \epsilon_{Z Z^{\prime}} r_{Z Z^{\prime}}^{2} Z^{\prime}_{\mu} J^{\mu}_{\rm neu} \,.
\end{equation}
Since the coupling of $Z^{\prime}$ to the neutral current is suppressed by $m_{Z^{\prime}}^{2} / m_{Z}^{2} \sim 10^{-8}$, its contribution to $Z^{\prime}$ phenomenology is negligible.
In the rest of this section, we discuss the observational constraints on $Z^{\prime}$ and $\varphi$.

\subsection{Experimental constraints}
\label{sec:exp-constraint}

\begin{figure}
  \centering
  \begin{minipage}{0.5\linewidth}
    \includegraphics[width=1.0\linewidth]{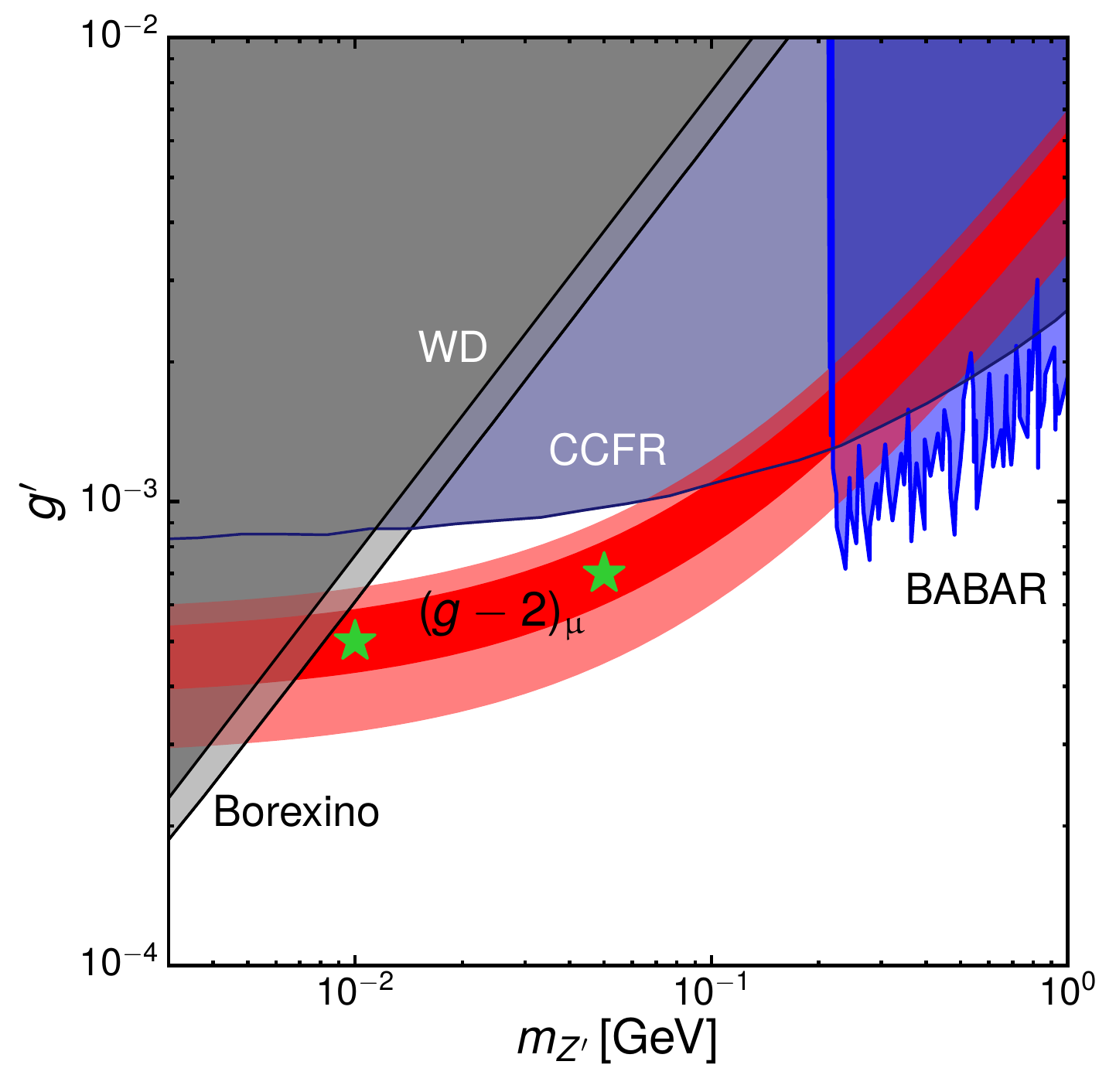}
  \end{minipage}%
  \begin{minipage}{0.5\linewidth}
    \includegraphics[width=0.94\linewidth]{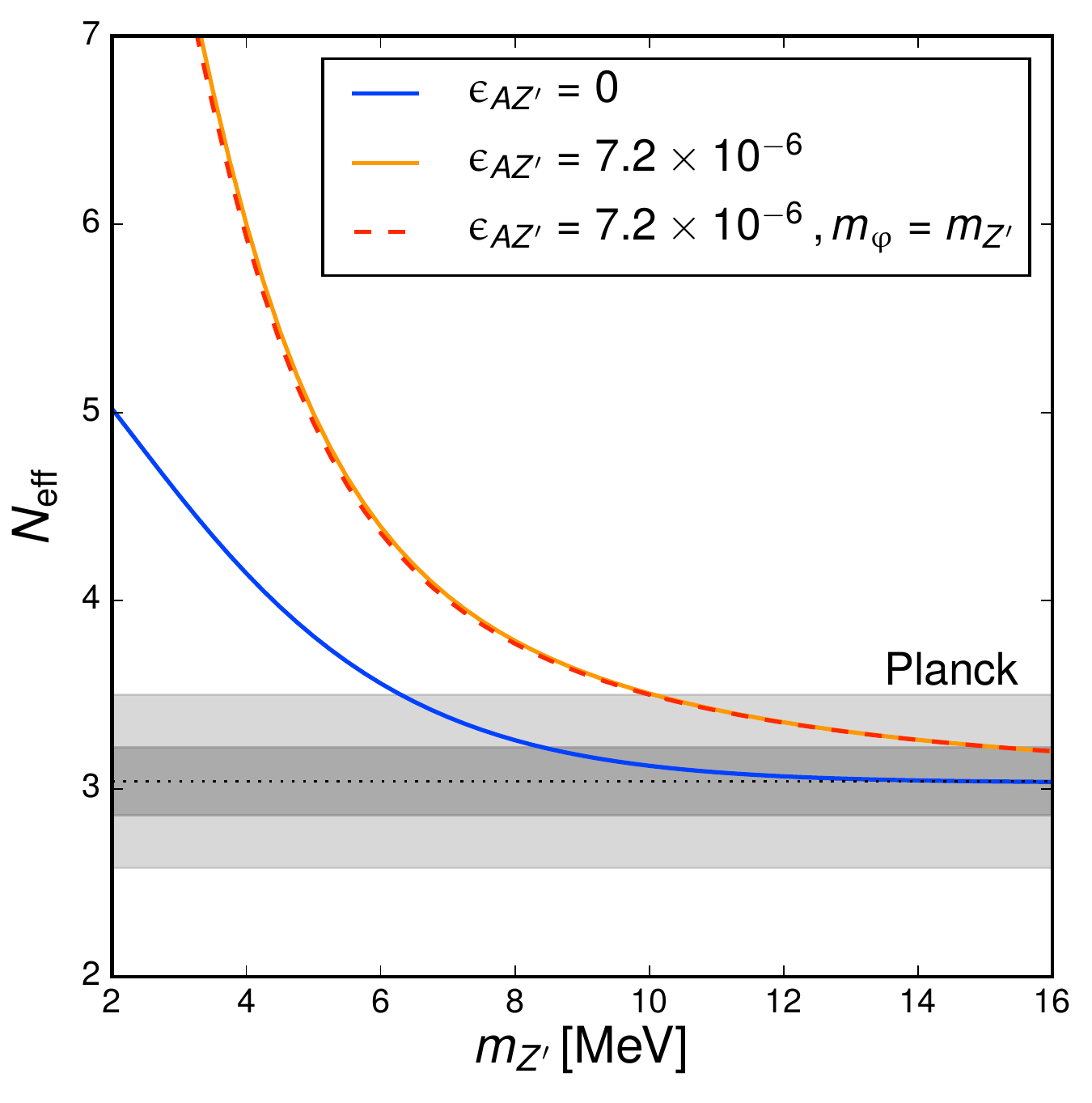}
  \end{minipage}
  \caption{Left: Parameter regions favored by the measurement of the muon anomalous magnetic moment within the $1 \sigma$ (red) and $2 \sigma$ (pink) limits, together with constraints (shaded) from various experiments, including BABAR (blue), CCFR (dark blue), Borexino (gray), and white dwarf cooling (dark gray). The green stars denote two examples shown in~Fig.~\ref{fig:si-xsect-scan}. Right: Predicted effective neutrino degrees of freedom at the temperature below $m_{e}$, for different $\epsilon_{AZ^{\prime}}$ values in the presence of only $Z'$ (solid), and both $Z'$ and $\varphi$ (dashed). The gray shaded regions denote the results from the Planck collaboration (\textit{Planck} TT, TE, EE+lowP+BAO) at the $1\sigma$ (dark) and $2\sigma$ (light) levels~\cite{Ade:2015xua}.}
  \label{fig:constraints}
\end{figure}

The muon anomalous magnetic moment $a_{\mu} = (g - 2)_{\mu} / 2$ is measured in the Brookhaven E821 experiment~\cite{Bennett:2006fi, Roberts:2010cj}, and its value shows a $3 \sigma$ deviation from the SM prediction.
Depending on the uncertainty in the hadronic vacuum polarization contributions, the difference is evaluated as $a_{\mu}^{\rm exp} - a_{\mu}^{\rm SM} = (26.1 \pm 8.0) \times 10^{-10}$~\cite{Hagiwara:2011af} or $a_{\mu}^{\rm exp} - a_{\mu}^{\rm SM} = (28.7 \pm 8.0) \times 10^{-10}$~\cite{Davier:2010nc}.%
\footnote{We adopt $\Delta a_{\mu}^{\rm LbL} = (10.5 \pm 2.6) \times 10^{-10}$~\cite{Prades:2009tw} as the hadronic light-by-light scattering contributions.
Another group evaluates it as $\Delta a_{\mu}^{\rm LbL} = (11.6 \pm 4.0) \times 10^{-10}$~\cite{Jegerlehner:2009ry}.
In any case, the discrepancy between the theoretical prediction and the experimental result is at the $3 \sigma$ level.}
Through its interaction with the muon, $Z^{\prime}$ provides a sizable correction to $(g - 2)_{\mu}$.
The 1-loop contribution is evaluated as~\cite{Baek:2001kca, Heeck:2011wj, Lynch:2001zr}
\begin{equation}
  \label{eq:zp-g-2}
  \Delta a_{\mu}^{Z^{\prime}}
  =
  \frac{g^{\prime 2}}{8 \pi^{2}}
  \int_{0}^{1} dx \frac{2m_{\mu}^{2} x^{2}(1 - x)}{x^{2} m_{\mu}^{2} + (1 - x) m_{Z^{\prime}}^{2}} \,.
\end{equation}
Since $\Delta a_{\mu}^{Z^{\prime}} \simeq g^{\prime 2} / (8 \pi^{2}) = 32 \times 10^{-10} \left( g^{\prime} / 5 \times 10^{-4} \right)^{2}$ for $m_{Z^{\prime}} \ll m_{\mu}$, the requirement of $\Delta a_{\mu}^{Z^{\prime}} < a_{\mu}^{\rm exp} - a_{\mu}^{\rm SM}$ gives $g^{\prime} \lesssim 5 \times 10^{-4}$.
Thus, the upper limit, $g^{\prime} \simeq 5\times 10^{-4}$, is favored for resolving the discrepancy.
In Fig.~\ref{fig:constraints} (left), we show the favored parameter regions for the $(g - 2)_{\mu}$ measurement, where we have used the result in Ref.~\cite{Hagiwara:2011af}.

Figure~\ref{fig:constraints} (left) also summarizes the current experimental constraints on the model, and we provide some details in the following:
\begin{itemize}
  \item
  {\bf Neutrino trident production.}
  The cross sections for $\nu N \to \nu N \mu {\bar \mu}$ at specific scattering energies have been measured in neutrino beam experiments, such as CHARM-II~\cite{Geiregat:1990gz} and CCFR~\cite{Mishra:1991bv}, and the results are in a good agreement with the SM predictions. We present the CCFR constraint on the model~\cite{Altmannshofer:2014pba} in Fig.~\ref{fig:constraints} (left). 
  \item
  {\bf $\Upsilon$ decays.}
  The BABAR experiment has searched for the $e {\bar e} \to \mu {\bar \mu} Z^{\prime}$ process followed by $Z^{\prime} \to \mu {\bar \mu}$ at the $\Upsilon$ resonance~\cite{TheBABAR:2016rlg}, resulting an upper limit on $g'$ for $200 \MeV \lesssim m_{Z^{\prime}} \lesssim 10 \GeV$.
  \item 
  {\bf $\nu \text{--} e$ scattering.}
  The Borexino experiment measures the interaction rate of the mono-energetic $862 \keV$ $^{7}$Be solar neutrino~\cite{Bellini:2011rx}. It puts a strong constraint on models that predict new $\nu \text{-} e$ interactions. 
 In this work, we calculate the $\nu \text{--} e$ reaction rate by taking into account the $Z^{\prime}$ contribution (see appendix~\ref{sec:borexino} for details). We require that the total reaction rate deviates from the SM prediction no more than 8\%~\cite{Bellini:2011rx} and obtain the bound.
  \item
  {\bf White dwarf (WD) cooling.}
  The white dwarf cooling also gives a strong limit~\cite{Bauer:2018onh}. The plasmon in the white dwarf star could decay to neutrinos through off-shell $Z^{\prime}$, and this process increases the cooling efficiency. In the SM, the effective operator $C_{V} G_{F} / \sqrt{2} \left[ {\bar \nu} \gamma^{\mu} (1 - \gamma^{5}) \nu \right] \left[ {\bar e} \gamma_{\mu} e \right]$, where $G_{F}$ is the Fermi constant and $C_{V} \simeq 0.964$~\cite{Dreiner:2013tja}, induces the plasmon decay through the electron loop. In our model, we have a similar operator, $G_{Z^{\prime}} \left[ {\bar \nu} \gamma^{\mu} (1 - \gamma^{5}) \nu \right] \left[ {\bar e} \gamma_{\mu} e \right]$, where $G_{Z^{\prime}} = - e g^{\prime} \epsilon_{A Z^{\prime}} /(2 m_{Z^{\prime}}^{2})$. Demanding $G_{Z^{\prime}} \lesssim C_{V} G_{F}$, we obtain $(g'/7.7\times10^{-4})^2(10~{\rm MeV}/m_{Z'})^2\lesssim1$, which is close to the Borexino bound.
\end{itemize}
It is remarkable that given these constraints there is still a viable parameter space for explaining $(g - 2)_{\mu}$, i.e., $5 \MeV \lesssim m_{Z^{\prime}} \lesssim 200 \MeV$ and $3.0 \times10^{-4} \lesssim g^{\prime} \lesssim 1.1 \times 10^{-3}$.
Since $Z^{\prime}$ does not couple to the electron directly, our model is much less constrained when compared to others invoking a new gauge boson, such as the U(1)$_{B - L}$ model~\cite{Harnik:2012ni} and the hidden photon model with the kinetic mixing (see, e.g., Ref.~\cite{Batley:2015lha}).
In what follows, we take two model examples: $(m_{Z^{\prime}}, g^{\prime}) = (10 \MeV, 5 \times 10^{-4})$ and $(50 \MeV, 7 \times 10^{-4})$ as indicated in Fig.~\ref{fig:constraints} (green stars).

\subsection{Cosmological constraint}
\label{sec:cosmo-constraint}

The main decay channel of $Z^{\prime}$ is $Z^{\prime} \to \nu {\bar \nu}$, and its lifetime is
\begin{align}
  \label{eq:lifetime-zp}
  \tau_{Z^{\prime}}
  &= 1 \times 10^{-14} \unit{sec} \bcm{g^{\prime}}{5 \times 10^{-4}}{-2} \bcm{m_{Z^{\prime}}}{10\MeV}{-1} \,.
\end{align}
For the model parameters that we are interested in, $\tau_{Z^{\prime}}$ is much shorter than the BBN timescale, $t \sim 1 \unit{sec} \, (1 \MeV / T)^{2}$.
Thus, $Z^{\prime}$ is in equilibrium with $\nu_{\mu}$ and $\nu_{\tau}$ through the decay and inverse decay, after neutrinos decouple from the SM plasma, $T = T_{\nu \text{-} {\rm dec}} \simeq 1.5 \MeV$.
Its energy density is suppressed by the Boltzmann factor at $T \ll m_{e}, m_{Z^{\prime}}$, and its direct contribution to the effective number of neutrino degrees of freedom $N_{\rm eff}$ is negligible. Note $N_{\rm eff}$ is related to the total energy density ($\rho_{\rm tot}$) and the photon density ($\rho_\gamma$) as
\begin{align}
\rho_{\rm tot} = \rho_{\gamma} \left[ 1 + N_{\rm eff} \frac{7}{8} \left( \frac{4}{11} \right)^{4/3} \right] \,.
\label{eq:deltangen}
\end{align}

Even in this case, the presence of $Z^{\prime}$ may modify $N_{\rm eff}$, since $Z^{\prime}$ injects energy only into $\nu_{\mu}$ and $\nu_{\tau}$ and it changes the ratio of the neutrino temperature to the photon temperature.
In the limit of $\epsilon_{A Z^{\prime}}$=0, we follow the analysis in Ref.~\cite{Kamada:2015era} to estimate the effect.
We assume that ($\gamma$, $e$), ($\nu_{e}$), and ($\nu_{\mu}$, $\nu_{\tau}$, $Z^{\prime}$) form independent thermal baths after the neutrino decoupling, and the comoving entropy density is conserved in each bath.
By following the temperature evolution in each sector, we evaluate $N_{\rm eff}$ at $T \ll m_{e}, m_{Z^{\prime}}$.
In Fig.~\ref{fig:constraints} (right), we show the lower bound on $m_{Z^{\prime}}$ (blue), i.e., $m_{Z^{\prime}} \gtrsim 6 \MeV$, where we take the upper limit $N_{\rm eff} < 3.5$~\cite{Ade:2015xua}.

In the presence of a sizable $\epsilon_{A Z^{\prime}}$ as in Eq.~\eqref{eq:azp-mix}, $Z^{\prime}$ can decay to $e {\bar e}$, and there is heat transfer between ($\nu_{\mu}, \nu_{\tau}, Z^{\prime}$) and ($\gamma, e$) baths.
In this case, we implement the heat transfer through $Z^{\prime} \to e {\bar e}$ in the evolution of the comoving entropy density as
\begin{align}
\frac{1}{a^{3}} \frac{d}{dt} [s_{\gamma} (T) a^{3} + 2 s_{e} (T) a^{3}] &= \frac{1}{T} \Gamma_{Z^{\prime} \to e {\bar e}} \,[\rho_{Z^{\prime}} (T^{\prime}) - \rho_{Z^{\prime}} (T)] \,,\label{eq:bbn-1} \\
\frac{1}{a^{3}} \frac{d}{dt} [2 s_{\nu_{\mu}} (T^{\prime}) a^{3} + 2 s_{\nu_{\tau}} (T^{\prime}) a^{3} + s_{Z^{\prime}} (T^{\prime}) a^{3}] &= - \frac{1}{T^\prime} \Gamma_{Z^{\prime} \to e {\bar e}} \,[\rho_{Z^{\prime}} (T^{\prime}) - \rho_{Z^{\prime}} (T)], \label{eq:bbn-2}\\
\frac{1}{a^{3}} \frac{d}{dt} [2 s_{\nu_{e}} (T_{\nu}) a^{3}] &= 0 \,, \label{eq:bbn-3}\;
\end{align}
where $T$, $T_\nu$, and $T'$, denote the temperatures of ($\gamma, e$), ($\nu_{e}$), and ($\nu_{\mu}, \nu_{\tau}, Z^{\prime}$), respectively, and $\Gamma_{Z^{\prime} \to e {\bar e}}$ is the decay width of $Z^{\prime} \to e {\bar e}$,
\begin{align}
  \label{eq:Zp-to-ee}
  \Gamma_{Z^{\prime} \to e {\bar e}}
  =
  \frac{\alpha\epsilon_{A Z^{\prime}}^{2} (m_{Z^{\prime}}^{2} + 2 m_{e}^{2}) \sqrt{m_{Z^{\prime}}^{2} - 4 m_{e}^{2}}}{3 m_{Z^{\prime}}^{2}} \,,
\end{align}
where $\alpha$ is the fine-structure constant and $m_{e}$ is the electron mass.
Note that $\rho_{Z^{\prime}}$ for the decay is evaluated with $T^{\prime}$, while $\rho_{Z^{\prime}}$ for the inverse decay is evaluated with $T$.
We numerically solve Eqs.~\eqref{eq:bbn-1}-\eqref{eq:bbn-3} with the Hubble expansion rate,
\begin{equation}
H = \sqrt{\frac{\rho_{\gamma} (T) + 2 \rho_{e} (T) + 2 \rho_{\nu_{e}} (T_{\nu}) + 2 \rho_{\nu_{\mu}} (T^{\prime}) + 2 \rho_{\nu_{\tau}} (T^{\prime}) + \rho_{Z^{\prime}} (T^{\prime})}{3 m_{\rm Pl}^{2}}},
\label{eq:Hubble}
\end{equation}
where $m_{\rm Pl}$ is the reduced Planck mass. See appendix~\ref{sec:temperature-evolution} for detailed discussion on the temperature evolution. We obtain $N_{\rm eff}$ as shown in Fig.~\ref{fig:constraints} (right). For $N_{\mathrm{eff}} < 3.5$ and $\epsilon_{A Z^{\prime}} \simeq 7.2 \times 10^{-6}$ (orange), corresponding to $g^{\prime} = 5 \times 10^{-4}$, the lower bound is $m_{Z^{\prime}} \gtrsim 10 \MeV$, and stronger than the limit for $\epsilon_{A Z^{\prime}} = 0$.

In this model, $\varphi \varphi \leftrightarrow Z^{\prime} Z^{\prime}$ and $\varphi \nu \leftrightarrow  Z^{\prime} \nu$ keep $\varphi$ in equilibrium with $Z^{\prime}$ and thus with SM particles below $T \sim 20 \GeV$ for $g^{\prime} = 5 \times 10^{-4}$.
To be conservative, we can assume that $m_{\varphi} > m_{Z^{\prime}}$ so that the lifetime of $\varphi$ is less than $1 \unit{s}$ (see appendix~\ref{sec:decay-width-varphi}).
In this case, the presence of $\varphi$ does not change the lower bound on $m_{Z^{\prime}}$ as shown in Fig.~\ref{fig:constraints} (right).
We comment on the possibility of $m_{\varphi} < m_{Z^{\prime}}$, where $\varphi$ decouples from the thermal bath when $\varphi \nu \leftrightarrow Z^{\prime} \nu$ becomes inefficient. In this case, $\varphi$ dominantly decays to four neutrinos long after the BBN. It can also decay to two neutrinos and two electrons with a small branching ratio $\sim e^{2} \epsilon_{A Z^{\prime}}^{2} / g^{\prime 2}$. The latter process could inject electromagnetic energy to the plasma and lead to observational consequences. If the $\varphi$ lifetime is longer than $10^{6} \, {\rm s}$, the energy injection could distort the CMB spectrum from the blackbody radiation (see, e.g., Ref.~\cite{Asaka:1999xd}). We find that $m_{\varphi}$ has a lower limit, i.e., $m_{\varphi} \gtrsim 6 \MeV$, to satisfy the COBE constraint~\cite{Fixsen:1996nj} for $(m_{Z^{\prime}}, g^{\prime}) = (10 \MeV, 5 \times 10^{-4})$, while $m_{\varphi} \gtrsim 20 \MeV$ for $(m_{Z^{\prime}}, g^{\prime}) = (50 \MeV, 7 \times 10^{-4})$. In addition, the observation of the light-element abundances can further tighten the constraint (see, e.g., Ref.~\cite{Kawasaki:2017bqm}). A detailed study of the cosmological constraint is beyond the scope of this work, and we will take the conservative assumption in the rest of the paper.

\section{Dark Matter phenomenology}
\label{sec:dmpheno}

In this section, we discuss DM phenomenology predicted in this model.
The mass matrix of the DM candidates $N$ and $\bar N$ can be diagonalized by a unitary matrix $U$ as
\begin{align}
  \label{eq:mass-mat}
  - {\cal L}_{\rm mass}
  =
  \frac{1}{2}
  \begin{pmatrix}
    N &\bar{N}
  \end{pmatrix}
        \begin{pmatrix}
          y_{N} \frac{v_{\Phi}}{\sqrt{2}} & m_{N} \\
          m_{N} & y_{\bar N} \frac{v_{\Phi}}{\sqrt{2}}
        \end{pmatrix}
                \begin{pmatrix}
                  N \\ \bar N 
                \end{pmatrix} 
    =
  \frac{1}{2}
  \begin{pmatrix}
    N_{1} & N_{2}
  \end{pmatrix}
        \begin{pmatrix}
          M_{1} & 0 \\
            0 & M_{2}
        \end{pmatrix}
                \begin{pmatrix}
                  N_{1} \\ N_{2}
                \end{pmatrix}  \,,
\end{align}
where $0 \leq M_{1} \leq M_{2}$ and
\begin{align}
  \label{eq:mass-eig}
  \begin{pmatrix}
    N_{1} \\ N_{2}
  \end{pmatrix}
  =
  U
  \begin{pmatrix}
    N \\ \bar N
  \end{pmatrix} \,.
\end{align}
Then the Lagrangian density in the 4-component spinor notation, where $N_{i}$ denotes a Majorana fermion, becomes
\begin{align}
  \label{eq:lag-eig-4comp}
  {\cal L}
  &\supset
    \frac{i}{2} \overline{N}_{i} \gamma^{\mu} \partial_{\mu} N_{i}
    - \frac{1}{2} M_{i} \overline{N}_{i} N_{i}
    + \frac{g^{\prime}}{4} Z^{\prime}_{\mu} \overline{N}_{i} \left[i \, {\rm Im} (U_{i 1} U_{j1}^{*} - U_{i 2} U_{j 2}^{*}) + {\rm Re}(U_{i 1} U_{j1}^{*} - U_{i 2}U_{j 2}^{*}) \, \gamma_{5} \right] \gamma^{\mu} N_{j}
    \notag\\
  &- \frac{1}{2 \sqrt{2}} \varphi \overline{N}_{i}
    \left[ {\rm Re}(y_{N} U_{i 1}^{*} U_{j 1}^{*} + y_{\bar N} U_{i 2}^{*} U_{j 2}^{*})
    - i \, {\rm Im}(y_{N} U_{i 1}^{*} U_{j 1}^{*} + y_{\bar N} U_{i 2}^{*} U_{j 2}^{*}) \gamma_{5} \right] N_{j} \,.
\end{align}

We consider two extreme cases to illustrate the main predictions of our model and further simplify the Lagrangian density.
\begin{itemize}
\item
{\bf Pseudo-Majorana DM}, where $|y_{\bar N}| v_{\Phi} > |y_{N}| v_{\Phi} \gg |m_{N}|$. After performing the phase rotation of $N$ and $\bar{N}$ such that both $y_N$ and $y_{\bar N}$ are positive, we have $M_{1} \approx y_{N} v_{\Phi} / \sqrt{2}$ and $M_2\approx y_{\bar{N}} v_{\Phi} / \sqrt{2}$. Note that DM phenomenology is unchanged for $|y_{N}| v_{\Phi} > |y_{\bar N}| v_{\Phi} \gg |m_{N}|$. In this limit, it turns out that we cannot simultaneously obtain the SIDM cross section, explain the $(g - 2)_{\mu}$ discrepancy, and realize viable cosmology.
We will not focus on this case in the rest of the paper.
\item
{\bf Pseudo-Dirac DM}, where $|m_{N}| \gg |y_{N}| v_{\Phi}, |y_{\bar N}| v_{\Phi}$. We choose a basis such that $m_N$ and $y^*_N+y_{\bar{N}}$ are positive and further restrict our discussion to the case where $y_{N} = y_{\bar N} \equiv y > 0$. In this limit, the DM sector respects $C_{L_{\mu} - L_{\tau}}$ ($N \leftrightarrow {\bar N}$ and $\Phi \to \Phi^{*}$) and the parity conjugation ($N \to i {\bar N}^{\dagger}$ and ${\bar N} \to i N^{\dagger}$).
The mass eigenstates are $N_{1} = (N - \bar{N}) / (\sqrt{2} i)$ and $N_{2} = (N + \bar{N}) / \sqrt{2}$, and the corresponding masses are $M_{1} \approx m_{N} - y v_{\Phi} / \sqrt{2}$ and $M_{2} \approx m_{N} + y v_{\Phi} / \sqrt{2}$.
The gauge and Yukawa interactions are
\begin{align}
  \label{eq:n1n2-yukawa-gauge}
  {\cal L}
  \supset
  -\frac{y}{2\sqrt 2}
  \varphi
  \left( - \overline{N}_{1} N_{1} + \overline{N}_{2}N_{2} \right)
  + i g^{\prime} Q_{N} Z_{\mu}^{\prime} \overline{N}_{2}\gamma^{\mu} N_{1} \,.
\end{align}
\end{itemize}
To explore DM phenomenology predicted in this pseudo-Dirac case, we need to specify five parameters: $m_{Z^{\prime}}$, $g^{\prime}$, $m_{\varphi}$, $M_{1}$, and $y$.
For the first two, we take the examples motivated by the $(g - 2)_\mu$ measurement as denoted by the green stars in Fig.~\ref{fig:constraints} (left).
We then scan the $(m_{\varphi}, M_{1})$ parameter space, while fixing $y$ by the relic density requirement accordingly, as discussed next.

\subsection{DM relic abundance}
\label{sec:dm-relic-abundance}

We determine the Yukawa coupling $y$ such that the $N_{1}$ relic density gives rise to the observed DM abundance.
We assume that $M_{1} \sim M_{2} \lesssim 100 \GeV$ so that the freeze-out occurs after the U$(1)_{L_{\mu} - L_{\tau}}$ symmetry breaking, $T < v_{\Phi} = \Order(10) \GeV$, and we can neglect the thermal corrections to $\Delta M = M_{2} - M_{1} = \sqrt{2} y v_{\Phi}$, $m_{\varphi}$, and $m_{Z^{\prime}}$.%
\footnote{This mass range is also compatible with the naturalness requirement of $\lambda_{\Phi}$.
The sizable DM self-scattering cross section requires $m_{\varphi}\sim 10 \MeV$, which corresponds to $\lambda_{\Phi} \sim 10^{-6}$ for $v_{\Phi} \sim10 \GeV$. The dominant correction to $\lambda_{\Phi}$ comes from the fermion box diagrams and is estimated as $y^{4} / (16 \pi^{2})$. For $y \sim10^{-2} \textup{--} 10^{-1}$, as determined by the relic density consideration, the correction to $\lambda_{\Phi}$ is less than a factor of $10^{-6}$ for $m_{N} \lesssim 100 \GeV$.}
Since $\Delta M = \Order(1) \GeV$, $N_{1}$ and $N_{2}$ are in equilibrium with each other during the freeze-out and we need to take into account their co-annihilation process~\cite{Griest:1990kh}. The annihilation channels are $N_{1} N_{1}, N_{2} N_{2} \to Z^{\prime} Z^{\prime}, \varphi \varphi$ and $N_{1} N_{2} \to \varphi Z^{\prime}$, and the corresponding thermally-averaged cross sections times relative velocity can be written as
\begin{align}
  \vev{\sigma_{\rm ann} v_{\rm rel}}_{11}
  =
  \vev{\sigma_{\rm ann} v_{\rm rel}}_{22}
  \simeq
  \frac{9 y^{4}}{64 \pi m_{N}^{2}} x^{-1},\;\;
    \vev{\sigma_{\rm ann} v_{\rm rel}}_{12}
  \simeq
  \frac{y^{4}}{64 \pi m_{N}^{2}} - \frac{9 y^{4}}{256 \pi m_{N}^{2}} x^{-1} \,,
  \label{eq:ann}
\end{align}
where $x = M_{1} / T$. In deriving Eq.~\eqref{eq:ann}, we have ignored the terms depending on $g'$, since they are too small to play a role. The effective annihilation cross section is 
\begin{align}
  \label{eq:eff-xsect}
  \vev{\sigma_{\rm ann} v_{\rm rel}}_{\rm eff}
  &=
    \vev{\sigma_{\rm ann} v_{\rm rel}}_{11} r_{1}^{2}
  +
    \vev{\sigma_{\rm ann} v_{\rm rel}}_{22} r_{2}^{2}
    +
    2 \vev{\sigma_{\rm ann} v_{\rm rel}}_{12} r_{1} r_{2} \,,
\end{align}
where $r_{1} = 1 / \left( 1 + e^{- \Delta x} \right)$, $r_{2} = e^{- \Delta x} / \left( 1 + e^{-\Delta x} \right)$, and $\Delta \equiv \Delta M / m_{N}$.
We equate this effective annihilation cross section to the canonical one, $\vev{\sigma_{\rm ann} v_{\rm rel}}_{\rm can} = 3 \times 10^{-26} \unit{cm^{3} / s}$, at $x_{f} = 20$ and derive the $y$ value for given ($m_\varphi$, $M_1$).
We have checked that the Sommerfeld effect is negligible for the freeze-out calculation (see, e.g., Refs.~\cite{Hisano:2006nn, Feng:2010zp}) for the DM mass range that we are interested in.

\subsection{DM self-scattering}
\label{sec:dm-self-scattering}
The elastic scattering of $N_{1}$ is dominated by the Yukawa interaction in Eq.~\eqref{eq:n1n2-yukawa-gauge}, which is represented by the following Yukawa potential for non-relativistic scattering:
\begin{align}
  \label{eq:yukawa-pot}
  V(r) = - \frac{y^{2}}{4\pi r}e^{-m_{\varphi} r} \,,
\end{align}
where the $y$ value is fixed by the relic density consideration as discussed above.
The scattering amplitude $f(\theta)$ is given by the partial wave expansion as 
\begin{align}
  \label{eq:partial-wave-amp}
  f(\theta) = \frac{2}{M_{1} v_{\rm rel}} \sum_{\ell = 0}^\infty (2 \ell + 1) e^{i \delta_{\ell}} \sin \delta_{\ell} P_{\ell} (\cos \theta) \,,
\end{align}
where $\delta_\ell$ is the phase shift and $P_{\ell}$ is the $\ell$th order Legendre polynomial.
We numerically calculate $\delta_\ell$ by solving the Schr\"odinger equation as in Ref.~\cite{Tulin:2013teo}.
Since $N_{1}$ is a Majorana fermion, the scattering particles are indistinguishable.
Then the differential cross section is given by the sum of spin-singlet and spin-triplet channel amplitudes as
\begin{align}
  \label{eq:diff-xsect-idnt}
  \frac{d\sigma}{d\Omega}
  &=
    \frac{1}{2} \times\left[
   \frac{1}{4}|f(\theta) + f(\pi - \theta)|^{2}
    +
    \frac{3}{4}|f(\theta) - f(\pi - \theta)|^{2}
    \right] \,,
\end{align}
where the initial state is assumed to be unpolarized~\cite{Kahlhoefer:2017umn}.

The standard cross section, $\sigma = \int d\Omega \, (d\sigma / d\Omega)$, is not an appropriate quantity for characterizing the effect of self-scattering on the structure formation, since the scattering through a light mediator receives strong enhancement at $\theta = 0$ and $\pi$, which do not affect the DM distribution.
Instead the transfer cross section is often used in the literature~\cite{Tulin:2017ara}, which is written for the scattering of identical particles as~\cite{Kahlhoefer:2013dca}
\begin{align}
  \label{eq:trans-xsect-indist}
  \sigma_{T}
  &=
  4 \pi \int_{0}^{1} d\cos \theta \, (1 - \cos \theta) \frac{d\sigma}{d\Omega} \,.
\end{align}

\begin{figure}
  \centering
  \begin{minipage}{0.5\linewidth}
    \includegraphics[width=1.0\linewidth]{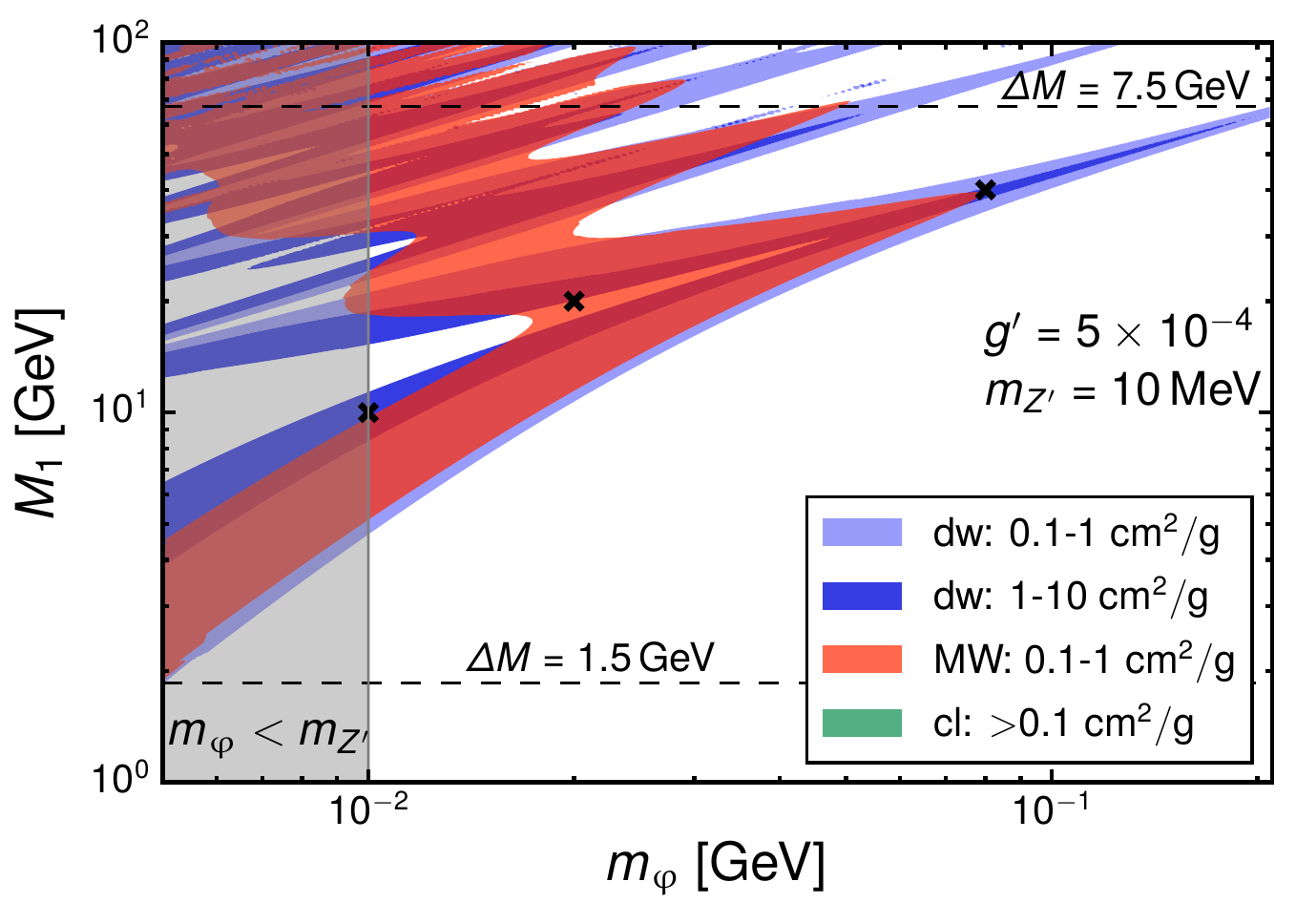}
  \end{minipage}%
  \begin{minipage}{0.5\linewidth}
    \includegraphics[width=1.0\linewidth]{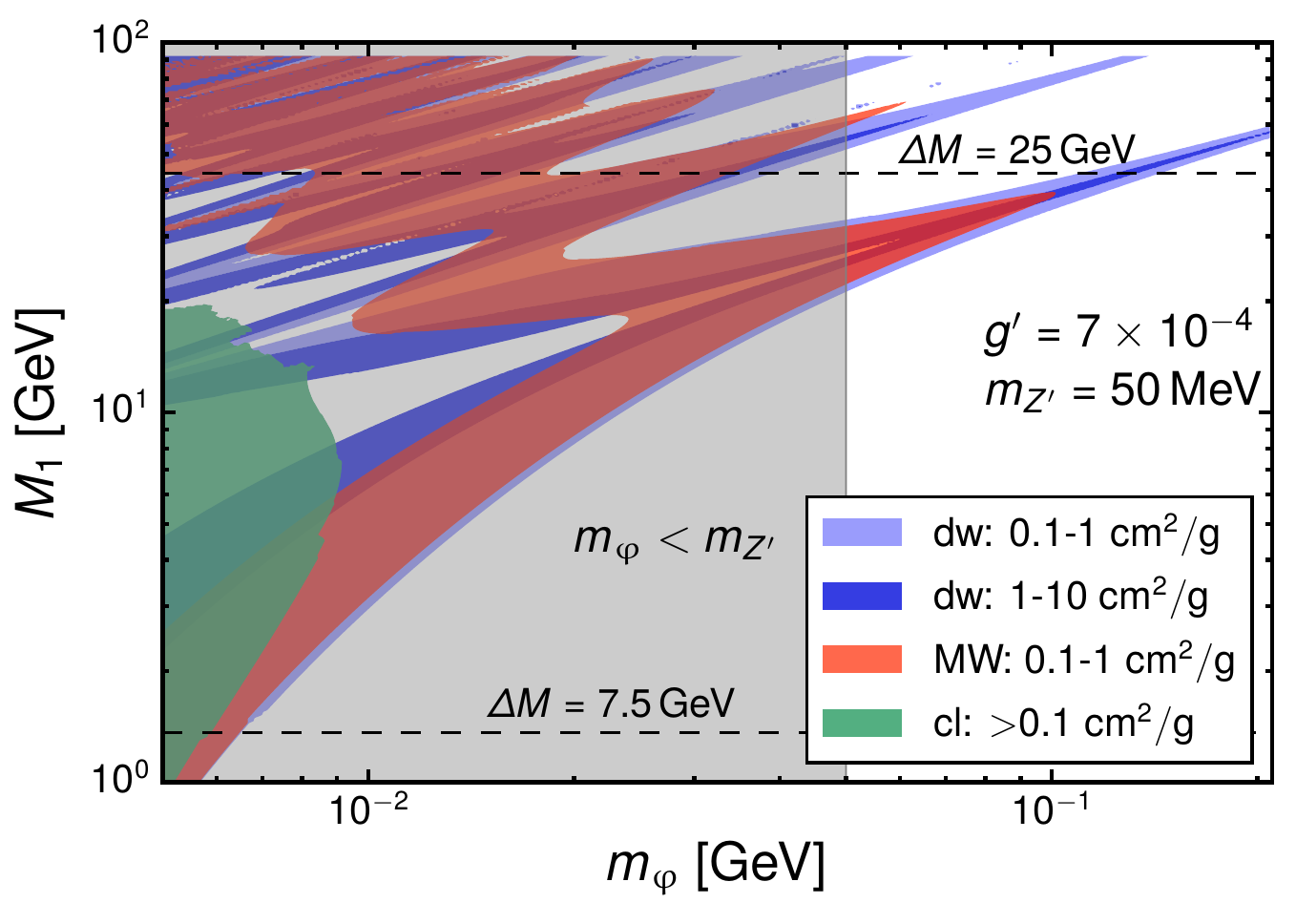}
  \end{minipage}
  \caption{SIDM parameter regions for dwarf galaxies (blue), Milky Way-size galaxies (red), and galaxy cluster (green).
    We take $(m_{Z^{\prime}}, g^{\prime}) = (10 \MeV, 5 \times 10^{-4})$ (left) and $(50 \MeV, 7 \times 10^{-4})$ (right).
    In the left panel, the self-scattering cross section is less than $0.1\unit{cm^2/g}$ in galaxy clusters.
  The BBN observations disfavor $m_{\varphi} < m_{Z^{\prime}}$ (gray). The dashed lines indicate the mass splittings between $N_1$ and $N_2$ states.
  The crosses (left) denote the benchmark cases shown in Fig.~\ref{fig:sigmav-pD}.}
  \label{fig:si-xsect-scan}
\end{figure}

In Fig.~\ref{fig:si-xsect-scan}, we show the parameter regions for $\sigma_{T} / M_{1} = 0.1 \text{--} 1 \unit{cm^{2} / g}$ (light blue) and $1\text{--}10 \unit {cm^{2} / g}$ (blue) in dwarf galaxies, $0.1\text{--}1 \unit{cm^{2} / g}$ (red) in Milky Way-size galaxies, and $> 0.1 \unit{cm^{2} / g}$ (green) in galaxy clusters, where we take $v_{\rm rel} = 30$, $200$, and $1000 \unit{km/s}$, respectively.
We have used the Born approximation when it is valid.
We see that with a light mediator, i.e., $m_{\varphi} = \Order(10) \MeV$, the SIDM cross section can be large in galaxies, while diminishing towards galaxy clusters, as shown in both panels.
However, when we impose the cosmological constraint (gray), i.e., $m_{\varphi} > m_{Z^{\prime}}$, most of the SIDM parameter space is excluded for the case of $g^{\prime} = 7 \times 10^{-4}$ and $m_{Z^\prime} = 50 \MeV$ (right).

\begin{figure}
  \centering
  \includegraphics[width=0.7\textwidth]{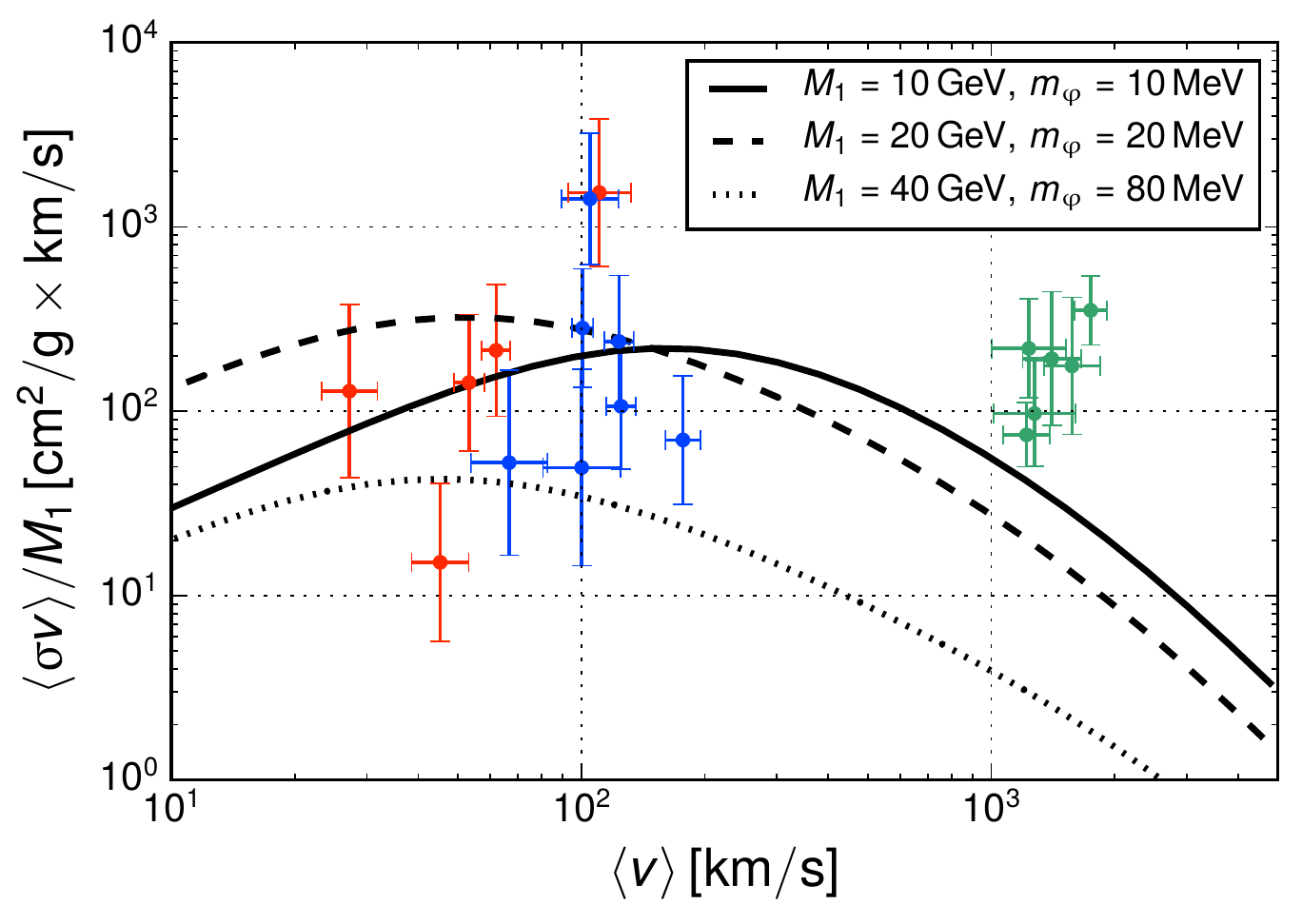}
  \caption{Self-interaction cross sections, $\vev{\sigma_{T} v} / M_{1}$, at the benchmark points taken from the left panel of Fig.~\ref{fig:si-xsect-scan}.
    We take $(M_{1}, m_{\varphi}) = (10 \GeV, 10 \MeV)$ (solid), $(20 \GeV, 20 \MeV)$ (dashed) and $(40\GeV, 80\MeV)$ (dotted).
    The points with errors are taken from Ref.~\cite{Kaplinghat:2015aga}, corresponding to dwarf galaxies (red), low surface brightness galaxies (blue), and galaxy clusters (green).}
  \label{fig:sigmav-pD}
\end{figure}

Figure~\ref{fig:sigmav-pD} shows the velocity-dependence of the self-scattering cross section $\vev{\sigma_{T} v_{\rm rel}} / M_{1}$, averaged over the Maxwell-Boltzmann distribution, for the three benchmark cases marked in Fig.~\ref{fig:si-xsect-scan} (left), together with the inferred values from stellar kinematics of dwarf (red) and low surface brightness (blue) galaxies, and galaxy clusters (green)~\cite{Kaplinghat:2015aga}. We see that at least two cases (solid and dashed) have a large self-scattering cross section to address the diversity problem of galactic rotation curves. On cluster scales, all three cases have a cross section below $\sim0.1~{\rm cm^2/g}$, as required by the cluster constraints~\cite{Kaplinghat:2015aga}. Interestingly, this upper limit is preferred if the DM self-interactions also explain the inferred density cores in the galaxy clusters~\cite{Newman:2012nw}. The case with $M_1=10~{\rm GeV}$ may explain observations of stellar kinematics in both galaxies and galaxy clusters.  

As indicated in Fig.~\ref{fig:si-xsect-scan}, all three benchmark cases are in the resonance regime, where the self-scattering cross section has a strong velocity dependence~\cite{Tulin:2012wi}, i.e., $\sigma_{T} / M_{1} \propto 1 / {v^{2}_{\rm rel}}$. Thus, the cross sections decrease from $\gtrsim 1 \unit{cm^{2} / g}$ in dwarf galaxies to $< 0.1 \unit{cm^{2} / g}$ in galaxy clusters. This is a generic feature of the model due to the tight constraints. Combining the muon $(g - 2)$ and the Borexino constraints, we require $m_{Z^{\prime}} \gtrsim 10 \MeV$. We further demand $m_{\varphi} \gtrsim m_{Z^{\prime}}$ so that $\varphi$ decays before the onset of the BBN, and the Yukawa coupling constant $y$ to be fixed by the DM relic abundance. After imposing all these constraints, we find that our SIDM model is in the resonant regime. 

\subsection{Direct and indirect searches}
\label{sec:direct-indir-search}
If the SIDM mediator decays through the Higgs portal, DM direct detection experiments have put a strong constraint on the mixing parameter $\lambda_{\Phi H}$~\cite{Kaplinghat:2013yxa, Kahlhoefer:2017umn, Ren:2018gyx} and the lifetime of the mediator is too long to be consistent with early Universe cosmology. In the model we consider, $\varphi$ decays to $Z^\prime$ to avoid over-closing the Universe. Thus, it does not need to mix with the SM Higgs boson to deplete its abundance. We set $\lambda_{\Phi H} = 0$ at the tree level. A finite value of $\lambda_{\Phi H}$ can be generated through a 2-loop process, but it is small and much below the experimental sensitivity. Moreover, the spin-dependent DM-nucleon scattering is inelastic, since the relevant interaction term involves two states, as shown in Eq.~\eqref{eq:n1n2-yukawa-gauge}. In this model, the predicted mass splitting is $\Delta M >1 \GeV$ (see Fig.~\ref{fig:si-xsect-scan}) which is much larger than the nuclei recoil energy in DM direct detection. Thus, DM direct searches do not constrain this model.  

For indirect detection, the annihilation processes, $N_{1} N_{1} \rightarrow Z^{\prime} Z^{\prime}$ and $\varphi\varphi$, are $p$-wave dominated, and their final states decay to the SM neutrinos. Since the upper bound on the DM annihilation cross section to the neutrinos is $\vev{\sigma_{\rm ann} v_{\rm rel}} \lesssim 10^{-24} \unit{cm^{3}/s}$~\cite{Frankiewicz:2015zma}, our model easily avoids the indirect detection constraints. 

Before closing this section, we comment on the scenario where $y_{N} \neq y_{\bar N}$. In this case, there is a pseudo-scalar interaction, $i \varphi N_{1} \gamma_{5} N_{1}$, in addition to the scalar one. This new term induces a DM-DM potential that is suppressed by a factor of $m_{\varphi}^{2}/m_{N_{1}}^{2}$, compared to the Yukawa one. But, it is strongly singular as $V(r) \propto 1 / r^{3}$. We expect it has a subdominant effect on DM self-interactions since in the Born approximation limit the pseudo-scalar cross section vanishes for $v_{\rm rel}\rightarrow0$. For the gauge interactions, we now need to include the DM axial current, $- (g^{\prime} \kappa / 2) Z_{\mu}^{\prime} \overline{N}_{1} \gamma_{5} \gamma^{\mu} N_{1}$, with $\kappa \sim y v_{\Phi} / m_{\rm N}$. It induces a DM-nucleon scattering cross section that decreases with the DM velocity. The current constraint, e.g., from the LUX experiment~\cite{Akerib:2013tjd} is satisfied if $|\kappa| < \Order(10^{2})$ for $g^{\prime} = 5 \times 10^{-4}$ with the momentum transfer $\sim 100 \MeV$~\cite{DEramo:2016gos}.

\section{Conclusion}
\label{sec:concl}

We have constructed a U$(1)_{L_{\mu} - L_{\tau}}$ model that could explain the $(g - 2)_{\mu}$ measurement and reconcile the discrepancies of CDM on galactic scales. In this model, the U$(1)_{L_{\mu} - L_{\tau}}$ Higgs boson acts as the light dark force carrier, mediating DM self-interactions. We thoroughly studied the constraints from the high-energy and intensity-frontier experiments and cosmological and astrophysical observations, and found a viable model parameter space. Since the mediator dominantly decays to the neutrinos, the model avoids tight constraints from DM indirect searches. To be consistent with the cosmological upper bound, $N_{\rm eff} < 3.5$, we found that the mediator $\varphi$ and gauge boson $Z^{\prime}$ masses satisfy the condition of $m_{\varphi} > m_{Z^{\prime}} \gtrsim 10 \MeV$.
Interestingly, with the same mass range of $Z'$, the model could also explain the dip feature~\cite{Aartsen:2014gkd, Aartsen:2017mau} in the IceCube neutrino spectrum~\cite{Araki:2014ona, Kamada:2015era, Araki:2015mya}.
Since the Higgs mixing term vanishes, our model does not lead to direct detection signals via the Higgs portal. In addition, our model naturally predicts a $\rm GeV$ mass splitting between two DM mass states, which forbids DM-nucleon scattering via the gauge boson-photon kinetic mixing.

Our model could be tested in future collider experiments, such as the facilities in the intensity frontier~\cite{Ibe:2016dir, Kaneta:2016uyt, Gninenko:2018tlp, Abdullah:2018, Kahn:2018cqs}. In addition, the observation of a nearby supernova may also provide important hints on the properties of $Z^{\prime}$, since the non-standard neutrino self-interaction shortens the mean free path and increases the neutrino diffusion time from the supernova core~\cite{Kamada:2015era}. On the model building aspect, we could introduce three right-handed neutrinos and realize the see-saw mechanism to reproduce the observed neutrino masses and mixings~\cite{Asai:2017ryy}.

\acknowledgements

The work of AK and KK was supported by IBS under the project code IBS-R018-D.
The work of KK was supported in part by the DOE grant DE-SC0011842 at the University of Minnesota.
The work of KY was supported by JSPS KAKENHI Grant Number JP18J10202.
HBY acknowledges support from U.S. Department of Energy under Grant No. DE-SC0008541 and the U.S. National Science Foundation under Grant No. NSF PHY-1748958 as part of the KITP  ``High Energy Physics at the Sensitivity Frontier" workshop.

\appendix

\section{Neutrino-electron scattering rate}
\label{sec:borexino}
We provide details on the constraint from the neutrino-electron scattering discussed in Sec.~\ref{sec:exp-constraint} (see also, e.g., Refs.~\cite{Bilmis:2015lja, Jeong:2015bbi}).
The electron-neutrino scattering cross sections are 
\begin{align}
  \frac{d \sigma_{\rm SM}(\nu_{e} e)}{dT} 
  &= \frac{G_F^{2}m_e}{2\pi}
  \left[
    \{(1 + g_{V}) \pm (1 + g_{A}) \}^{2} + \{(1 + g_{V}) \mp (1 + g_{A}) \}^{2} \left(1 - \frac{T}{E_{\nu}} \right)^{2} 
  \right. \nonumber \\
  &
  \left.
    - (g_{V} - g_{A})(g_{V} + g_{A} + 2) \frac{m_{e} T}{E_{\nu}^{2}}
  \right] \,, \\
  \frac{d \sigma_{\rm SM}(\nu_{\alpha} e)}{dT} 
  &= \frac{G_{F}^{2} m_{e}}{2 \pi}
  \left[
    \{(g_{V} \pm g_{A}) + (g_{V} \mp g_{A}) \}^{2} \times \left(1 - \frac{T}{E_\nu} \right)^{2} 
    - (g_{V}^{2} - g_{A}^{2}) \frac{m_{e} T}{E_{\nu}^{2}}
  \right] \,, \\
  \frac{d \sigma_{Z^{\prime}}(\nu_{\alpha} e)}{dT} 
  &= \frac{g^{\prime 2}(e \epsilon_{A Z^{\prime}})^{2} m_{e}}{4 \pi (2 m_{e} T + m_{Z^{\prime}}^{2})^{2}}
  \left[
    1 + \left(1 - \frac{T}{E_\nu}\right)^{2} - \frac{m_{e} T}{E_{\nu}^{2}}
  \right] \nonumber \\
  &+ \frac{g^{\prime} e \epsilon_{A Z^{\prime}} G_{F} m_{e}}{\sqrt{2} \pi (2 m_{e} T + m_{Z^{\prime}}^{2})}
  \left[
    (g_{V} + g_{A}) + (g_{V} - g_{A}) \left(1 - \frac{T}{E_{\nu}}\right)^{2} - g_{V} \frac{m_{e} T}{E_{\nu}^{2}}
  \right] \,,
\end{align}
where $g_{V} = -1/2 + 2 s_{W}^{2}$ and $g_{A} = -1/2$.
The upper (lower) sign corresponds to the neutrino (anti-neutrino) and $\alpha = \mu, \tau$.
The incident neutrino energy is $E_\nu = 862 \keV$ for the $^{7}$Be solar neutrino.
$T$ is the electron recoil energy and its maximal value is $T_{\rm max} = 2 E_{\nu}^{2} / (m_{e} + 2 E_{\nu})$.
The minimal value of $T$ is $T_{\rm min} \simeq 270 \keV$ for the Borexino experiment~\cite{Bellini:2011rx, Bilmis:2015lja}.
Then, the total reaction rate $R_{\rm tot}$ is given by
\begin{align}
 \frac{dR_{\rm tot}}{dT} = t_{\rm exp} \, \rho_{e} \int dE_\nu\frac{d\sigma_{\rm tot}}{dT}\frac{d\Phi_{\nu_e}}{dE_\nu},
\label{eq:Rtot}
\end{align}
where $t_{\rm exp}$ and $\rho_{e}$ are the exposure time and the electron number density of the target, respectively.
We assume that the neutrino spectrum $d\Phi_{\nu_{e}} / dE_{\nu}$ is mono-energetic at $E_{\nu} = 862 \keV$. Since we are only interested in the ratio between the total reaction rates with and without $Z^{\prime}$, the overall normalization is irrelevant for our discussion.
We define the total cross section used in Eq.~\eqref{eq:Rtot} as
\begin{align}
  \sigma_{\rm tot}(\nu e) = P_{e e} \, \sigma_{\rm SM}(\nu_{e} e) + (1 - P_{e e}) \sum_{\alpha} \left[ \sigma_{\rm SM}(\nu_{\alpha} e) + \sigma_{Z^{\prime}}(\nu_{\alpha} e) \right] \,,
\end{align}
with the electron neutrino survival probability $P_{e e} \simeq 0.51$~\cite{Bellini:2011rx}.
For the SM prediction, one drops the $\sigma_{Z^{\prime}}$ contribution.

\section{Temperature evolution around the BBN}
\label{sec:temperature-evolution}
We summarize the basic thermal quantities and show how the temperatures evolve around the BBN.
The thermal distribution of particle $a$ takes a form of
\begin{align}
\label{eq:distribution}
f_{a} (p, T) = \frac{1}{e^{\sqrt{p^{2} + m_{a}^{2}} / T} \mp 1} \,,
\end{align}
where $-$ ($+$) is taken when particle $a$ is boson (fermion).
The entropy and energy densities of particle $a$ are given by
\begin{align}
s_{a} (T) &= g_{a} \int^{\infty}_{0} \frac{4 \pi p^{2} dp}{(2 \pi)^{3}} \left[ \frac{\sqrt{p^{2} + m_{a}^{2}}}{T} + \frac{p^{2}}{3 T \sqrt{p^{2}+m_{a}^{2}}} \right] f_{a} (p, T) \,, \\
\rho_{a} (T) &= g_{a} \int^{\infty}_{0} \frac{4 \pi p^{2} dp}{(2 \pi)^{3}} \sqrt{p^{2} + m_{a}^{2}} \, f_{a} (p, T) \,,
\end{align}
respectively, where $g_{a}$ is the spin degrees of freedom of particle $a$.
When $m_{a} \ll T$, the above integrations can be done analytically and read as
\begin{align}
s_{a} (T) &= g_{a} \left( \times \frac{7}{8} \right) \frac{2 \pi^{2}}{45}  T^{3} \,, \\
\rho_{a} (T) &= g_{a} \left( \times \frac{7}{8} \right) \frac{\pi^{2}}{30} T^{4} \,,
\end{align}
where the factor of $7/8$ is taken when particle $a$ is fermion.

\begin{figure}[t]
  \centering
  \begin{minipage}{0.5\linewidth}
    \includegraphics[width=0.97\linewidth]{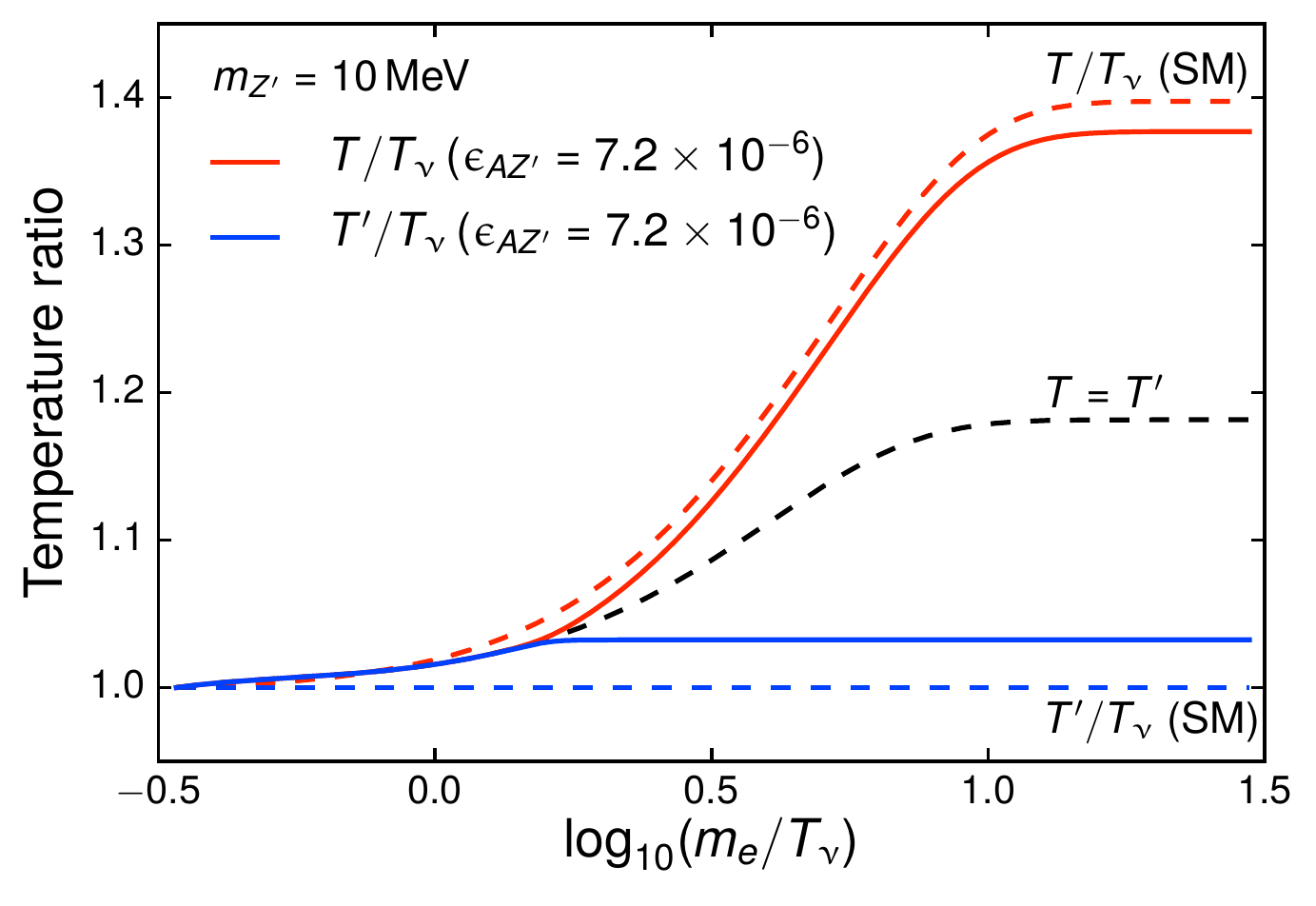}
  \end{minipage}%
  \begin{minipage}{0.5\linewidth}
    \includegraphics[width=1.0\linewidth]{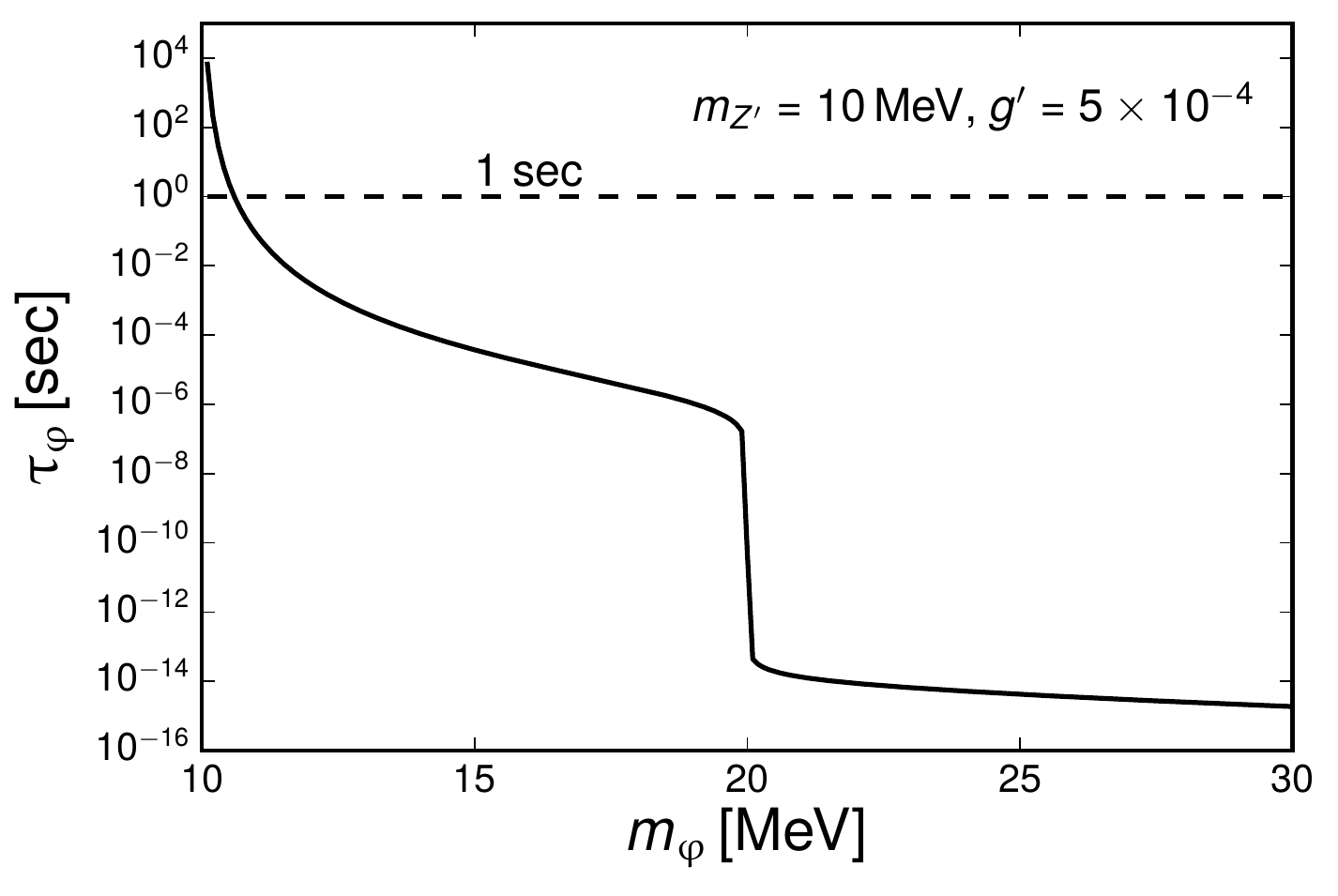}
  \end{minipage}
  \caption{Left: Evolution of $T / T_{\nu}$ (red) and $T^{\prime} / T_{\nu}$ (blue) for $\epsilon_{A Z^{\prime}} = 7.2 \times 10^{-6}$ (solid) and $0$ (dashed).
  Shown also is the case for the tightly coupled limit, i.e., $T = T^{\prime}$ (black dashed).
  Right: Lifetime of $\varphi$ as a function of $m_{\varphi}$.}
  \label{fig:appendix}
\end{figure}

In Sec.~\ref{sec:cosmo-constraint}, we introduced three temperatures: $T$ is the temperature of ($\gamma$, $e$), $T_{\nu}$ is that of ($\nu_{e}$), and $T^{\prime}$ is that of ($\nu_{\mu}$, $\nu_{\tau}$, $Z^{\prime}$).
We follow Eqs.~\eqref{eq:bbn-1}-\eqref{eq:bbn-3} and Eq.~\eqref{eq:Hubble}.
In Fig.~\ref{fig:appendix} (left), we show the evolution of $T / T_{\nu}$ (red) and $T^{\prime} / T_{\nu}$ (blue) for $\epsilon_{A Z^{\prime}} = 7.2 \times 10^{-6}$ (solid) and the SM (dashed), together with the case where ($\gamma$, $e$, $\nu_{\mu}$, $\nu_{\tau}$, $Z^{\prime}$) forms one thermal bath and thus $T = T^{\prime}$ (black dashed).
Note that $T_{\nu}$ is not affected by the presence of $Z^{\prime}$.
We see that $Z^{\prime} \rightarrow e {\bar e}$ lowers $T$, while raises $T^{\prime}$, when compared to the case of the SM.
This is because heat from $e {\bar e}$ annihilation is partially transferred from ($\gamma$, $e$) to ($\nu_{\mu}$, $\nu_{\tau}$, $Z^{\prime}$).

\section{Decay width of $\varphi$}
\label{sec:decay-width-varphi}
For $m_{\varphi} \geq 2 m_{Z^{\prime}}$, the dominant decay channel is $\varphi \to Z^{\prime} Z^{\prime}$ and the decay width is 
  \begin{align}
    \label{eq:phi-to-zpzp}
    \Gamma_{\varphi\to Z^{\prime} Z^{\prime}}
    &=
      \frac{g^{\prime 2} Q_{\Phi}^{2}}{32 \pi}
      \frac{\sqrt{m_{\varphi}^{2} - 4 m_{Z^{\prime}}^{2}}(m_{\varphi}^{4} - 4 m_{\varphi}^{2} m_{Z^{\prime}}^{2} + 12 m_{Z^{\prime}}^{4})}{m_{\varphi}^{2} m_{Z^{\prime}}^{2}} \,.
  \end{align}
  
While for $m_{Z^{\prime}} \leq m_{\varphi} < 2 m_{Z^{\prime}}$, the 3-body decay process, i.e., $\varphi \to Z^{\prime} \nu {\bar \nu}$, dominates and the width is given by
  \begin{align}
    \label{eq:width-phi-zpnunu}
    \Gamma_{\varphi \to Z^{\prime} \nu \bar{\nu}}
    &=
      \frac{g^{\prime 4} Q_{\Phi}^{2} m_{Z^{\prime}}^{4}}{32 \pi^{3} m_{\varphi}^{3}}
      F\left( \frac{m_{\varphi}}{m_{Z^{\prime}}} \right) \,,
  \end{align}
  where
  \begin{align}
    \label{eq:F}
    F(r)
    & =
     \int_{0}^{(r - 1)^{2}} dx \int_{y_{\rm min}}^{y_{\rm max}} dy
      \frac{r^{2}(y - 1) - x (y - 2) - y (y - 1)}{(x - 1)^{2} + w^{2}} \,,
    \\
    y_{\rm max} &= \frac{r^{2} + \sqrt{r^{4} - 2 r^{2} (x + 1) + (x - 1)^{2}} - x + 1}{2} \,,
    \\
    y_{\rm min} &= \frac{r^{2} - \sqrt{r^{4} - 2 r^{2} (x + 1) + (x - 1)^{2}} - x + 1}{2} \,,
  \end{align}
In Fig.~\ref{fig:appendix} (right), we show the lifetime of $\varphi$ vs its mass. For $m_{\varphi} > m_{Z^{\prime}}$, $\tau_{\varphi}$ is less than $1 \unit{s}$.



\bibliographystyle{utphys}
\bibliography{SIDM_Lmu-Ltau}

\end{document}